\documentclass{article}

\PassOptionsToPackage{numbers,sort&compress}{natbib}
\usepackage[preprint]{neurips_2026}

\usepackage{graphicx}
\usepackage[utf8]{inputenc}
\usepackage[T1]{fontenc}
\usepackage{hyperref}
\usepackage{url}
\usepackage{booktabs}
\usepackage{makecell}
\usepackage{amsfonts}
\usepackage{nicefrac}
\usepackage{microtype}
\usepackage{xcolor}
\usepackage{amsmath}

\newcommand{\addref}[1][\relax]{%
  \ifx#1\relax
    {\color{red} [cite]}%
  \else
    {\color{red} [cite: #1]}%
  \fi
}

\title{POPSICLE: Benchmark Datasets for Segmentation and Localization in CryoET}

\author{%
  \textbf{Jonathan Schwartz}$^1$, \quad
  \textbf{Utz Heinrich Ermel}$^1$, \quad
  \textbf{C.~Braxton Owens}$^2$, \quad \\
  \textbf{Zhuowen Zhao}$^1$, \quad
  \textbf{Ariana Peck}$^1$, \quad
  \textbf{Gus L.W. Hart}$^2$, \quad \\
  \textbf{Grant J. Jensen}$^2$, \quad
  \textbf{Bridget Carragher}$^1$, \quad
  \textbf{Dari Kimanius}$^{1,*}$ \\
  \\
  $^1$Biohub, Redwood City, CA 94063, USA \\
  $^2$Brigham Young University, Provo, UT, 84602, USA \\
  $^*$Corresponding author: \texttt{dari.kimanius@biohub.org}
}

\begin{document}

\maketitle

\begin{abstract}
Cryo–electron tomography (cryoET) has emerged as a powerful tool in structural and cellular biology by enabling direct visualization of macromolecular structures within intact cells, thereby linking molecular architecture to cellular organization in a native context. Realizing the full potential of cryoET, however, increasingly depends on advances in computational analysis, particularly machine learning (ML), to interpret its complex and information-rich data. Despite rapid progress, ML development for cryoET remains bottlenecked by the lack of standardized, well-annotated benchmarks. Existing evaluations are typically small, task-specific, and are assembled in isolation, limiting robust comparisons across methods. 
Here, we present POPSICLE, a benchmark suite for cryoET segmentation and macromolecular localization built from the CryoET Data Portal—an open, ML-ready repository of tomographic data, metadata, and annotations. POPSICLE spans eukaryotic and prokaryotic systems, both purified and fully \textit{in situ} samples, and dense voxel-wise segmentation as well as sparse localization tasks. Built on a living data resource, it can expand as new datasets and annotations become available. Baseline experiments reveal substantial variation in model rankings across tasks, underscoring the need for benchmarks tailored to the unique characteristics of cryoET rather than evaluation practices adapted from adjacent biomedical imaging domains. POPSICLE thus provides an open and extensible foundation for reproducible ML evaluation in cryoET.
\end{abstract}

\section{Introduction}
\label{sec:intro}
\begin{figure}
\centering
\includegraphics[width=\linewidth]{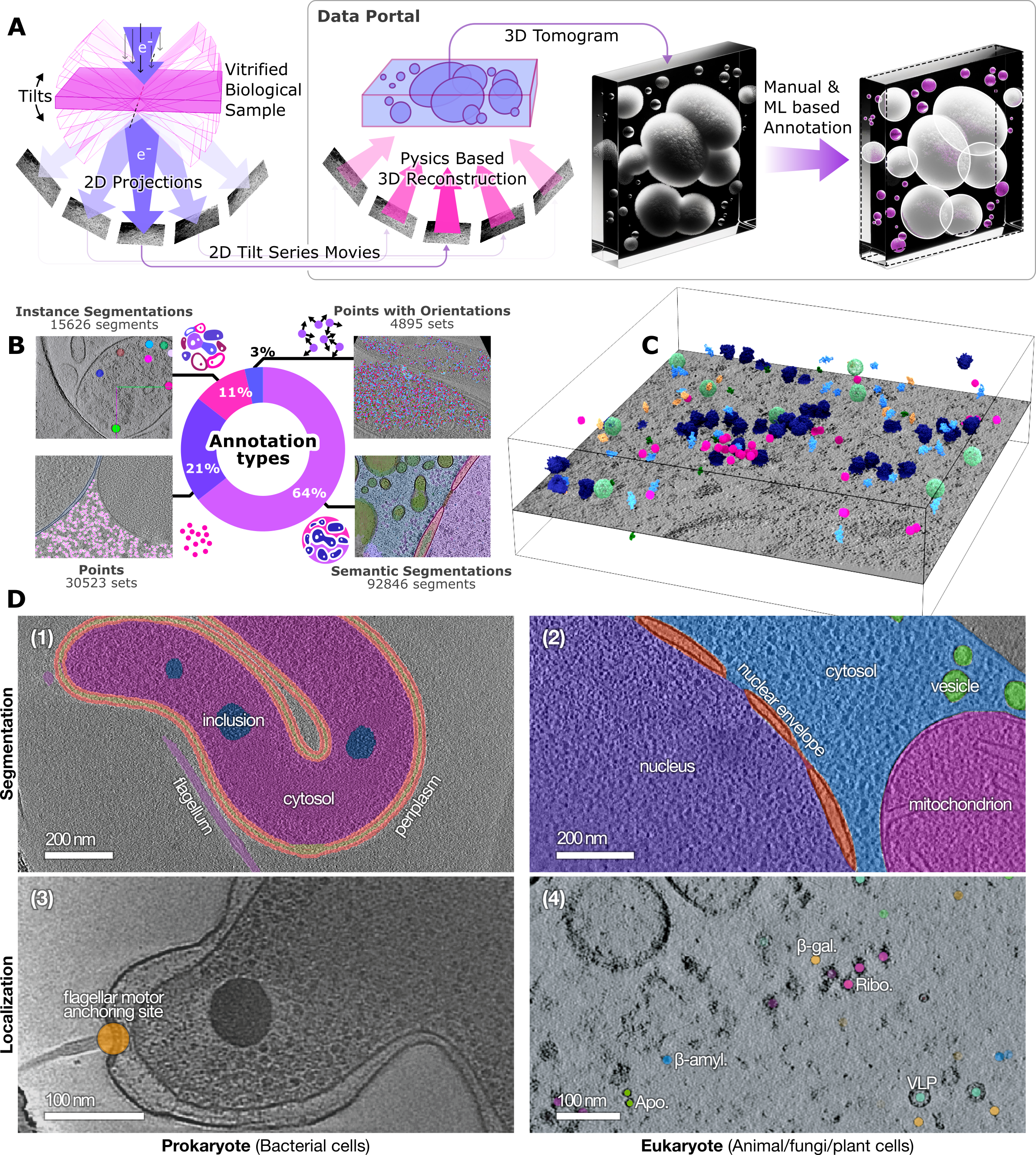}
\caption{
\textbf{Overview of data structures, processing, and annotations.}
\textbf{(A)} Schematic of the cryoET processing pipeline, from tilt-series acquisition in the microscope to tomographic 3D reconstruction and downstream annotation.
\textbf{(B)} Schematic of the four core data types available through the CryoET Data Portal.
\textbf{(C)} Three-dimensional visualization of localized molecular targets overlaid on a tomogram slice.
\textbf{(D)} Representative examples of the datasets and annotation modalities included in POPSICLE: (1) segmentation of a whole bacterial cell; (2) segmentation of subcellular compartments in a yeast cell; (3) localization of the bacterial flagellar motor anchoring site; and (4) localization of multiple molecular targets in a cell-like phantom dataset.
}
\label{fig:overview}
\end{figure}

Cryo-electron tomography (cryoET) enables three-dimensional (3D) imaging of biological specimens in their near-native state at nanometer-scale resolution~\cite{Beck2016} (Fig.~\ref{fig:overview}A). Unlike methods that isolate molecules from their cellular context, cryoET can resolve macromolecular structures directly inside intact cells, making it a powerful tool for studying how molecular architecture gives rise to cellular function~\cite{hutchings2018fine}. This ability to connect molecular structure with spatial organization has made cryoET increasingly important in structural and cellular biology~\cite{young2023bringing}.

As cryoET datasets grow in size and complexity, machine learning (ML) is needed to automate analysis that is otherwise labor-intensive, difficult to standardize, and hard to scale. CryoET poses a challenging regime for ML: tomograms are noisy, anisotropic, and shaped by acquisition artifacts such as restricted tilt ranges and reconstruction-dependent distortions~\cite{yan2019mbir, aretomo3}. The structures of interest also span multiple spatial scales, from cellular organization to individual macromolecular complexes, and vary widely across specimens, organisms, and imaging conditions~\cite{hutchings2018fine}. These properties make generalization difficult even when ML models perform well on a single dataset.

Evaluation in cryoET is further complicated by the structure of the tasks themselves. Dense segmentation is needed to recover membranes, organelles, and other cellular compartments, whereas sparse localization is needed to identify individual molecular complexes (Fig.~\ref{fig:overview}C) for downstream structural analysis~\cite{last2024streamlining, moebel2021deep}. These tasks differ in supervision, output representation, and evaluation, so success in one does not imply success in the other. At the same time, expert annotations are scarce and costly to produce, making most existing datasets small, narrow in scope, and insufficient for robust benchmarking~\cite{peck2025kaggle}.

As a result, current evaluation practices in cryoET remain fragmented. Datasets are often assembled for a single task under a single imaging condition with dataset-specific preprocessing and annotation conventions~\cite{peck2025kaggle, motorbench2025biorxiv}. Models are therefore commonly evaluated on in-distribution test sets that overstate real-world performance, while differences in splits, processing pipelines, and label quality make comparisons across methods unreliable~\cite{touchstone_benchmark}. More fundamentally, cryoET analysis often requires reasoning across scales, from cellular compartments to individual macromolecular targets, so evaluation limited to a single task gives only a partial view of model capability~\cite{schiotz2024serial, hutchings2018fine}. For this reason, we treat segmentation and localization as complementary benchmark tasks rather than isolated problems. Segmentation captures cellular context, whereas localization captures discrete molecular targets within that context. Evaluating them together can test whether models generalize across the main prediction regimes in cryoET and provide a broader foundation for future multi-task learning.

We address this gap with \textbf{POPSICLE} (\textbf{P}article/\textbf{O}bject \textbf{P}icking \& \textbf{S}egmentation \textbf{I}n \textbf{C}ryoET \textbf{L}earning \& \textbf{E}valuation), a unified benchmark for cryoET segmentation and macromolecular localization. POPSICLE is built on the CryoET Data Portal, an open, ML-ready resource that provides standardized tomographic data, annotations, metadata, and programmatic access~\cite{ermel2024data} (Fig.~\ref{fig:overview}A,B and Fig.~\ref{fig:numbers}; Appendix~\ref{sec:dataset_overview} and \ref{sec:data_structure}). Because the portal is a living database that continuously grows through new data deposits, annotations, and community contributions, POPSICLE is designed to grow with it as the underlying data resource evolves (Fig.~\ref{fig:numbers}D). 

POPSICLE brings together datasets spanning multiple organisms, imaging conditions, spatial scales, and annotation modalities within a single evaluation framework, enabling direct comparison of methods across the main task regimes at this processing stage in cryoET analysis~\cite{deepict, last2024streamlining}. By standardizing data organization, task definitions, and evaluation procedures, POPSICLE provides a reproducible alternative to prior task-specific evaluations. Furthermore, we make POPSICLE accessible through the copick toolkit~\cite{copick}, which provides a unified interface to tomograms, dense segmentations, and point annotations across local, shared, and cloud-backed storage, and interoperates with tools such as ChimeraX~\cite{chimerax} and Napari~\cite{sofroniew_2026_20058236} for visualization and curation (Appendix~\ref{sec:tools}). 

We benchmark representative convolutional, transformer-based, and cryoET-specific architectures and find that no single model is strong across tasks. Performance differs substantially between segmentation and localization, showing that evaluation practice and model architectures from adjacent domains such as medical imaging cannot be transferred without adaptation~\cite{touchstone_benchmark}.

Our main contributions are summarized as follows:
\begin{itemize}
    \item \textbf{A unified cryoET benchmark across tasks and regimes:} POPSICLE evaluates both dense compartment segmentation and sparse macromolecular localization across eukaryotic and prokaryotic systems, controlled and fully \textit{in situ} samples, and multiple biological and experimental settings. In total, the benchmark comprises 2,993 annotated tomograms.
    \item \textbf{Expanded benchmark coverage through new annotations:} We add new dense annotations for the bacterial segmentation dataset, extending the range of cellular-scale cryoET evaluation supported by the benchmark.
    \item \textbf{Integration with the CryoET Data Portal and copick:} We connect POPSICLE directly to the CryoET Data Portal through the copick toolkit, providing reproducible access to tomograms, annotations, dataset splits, and associated metadata.
    \item \textbf{Consolidated community reference results:} We incorporate challenge-derived reference results from two public Kaggle competitions, preserving strong community baselines for both multi-class and single-class localization tasks.
    \item \textbf{A comparative evaluation across model families:} We train and evaluate representative convolutional, transformer-based, and cryoET-specific architectures under a shared protocol for both segmentation and localization, and show that model rankings are strongly task- and dataset-dependent.
\end{itemize}

\section{Related Work}
\label{related_work}

Recent work in adjacent domains has shown that weak benchmark design can distort conclusions: small test sets, predominantly in-distribution evaluation, and inconsistent protocols can produce rankings that do not reflect real-world performance~\cite{touchstone_benchmark}. These concerns are especially important in cryoET, where annotated data are limited, tasks are heterogeneous, and evaluation is still largely organized around isolated datasets and study-specific protocols~\cite{last2024streamlining, moebel2021deep, lamm2022membrain}.

Existing cryoET benchmarks have focused primarily on macromolecular localization and classification. The SHREC cryoET challenges established standardized evaluation settings for particle detection and classification, but relied on simulated tomograms rather than experimental data, limiting realism with respect to the artifacts, heterogeneity, and annotation ambiguity of real cryoET volumes~\cite{gubins2020shrec}. More recent efforts have moved toward realistic experimental benchmarks. For example, the Phantom dataset introduced a large experimentally acquired resource with expert annotations for molecular localization~\cite{peck2025kaggle}, and MotorBench introduced an expert-annotated benchmark for bacterial flagellar motor localization in cellular tomograms~\cite{motorbench2025biorxiv}. These datasets substantially improve realism, but they remain task-specific and are centered on sparse localization.

Segmentation is equally central to cryoET analysis, with recent work emphasizing its importance for recovering cellular organization and supporting downstream biological interpretation. Unfortunately, current segmentation efforts are driven mainly by methods and software pipelines rather than by broadly adopted benchmark datasets or standardized evaluation protocols~\cite{rice2023tomotwin, guoule2024deepetpicker, last2024streamlining}. As a result, cryoET still lacks a unified benchmark that supports both dense voxel-wise segmentation and sparse localization within a common evaluation framework.

POPSICLE addresses this gap by moving cryoET benchmarking away from one-off challenge releases and toward a single, unified framework that can incorporate new annotation tasks, biological targets, and datasets over time. Anchoring the benchmark in the CryoET Data Portal makes this continual extension practical within a shared data and metadata infrastructure~\cite{ermel2024data}.

\section{Benchmark Tasks}

\label{sec:tasks}

POPSICLE covers the two main annotation regimes in cryoET: dense voxel-wise segmentation of cellular structure and sparse localization of macromolecular targets  (Fig.~\ref{fig:overview}B,D). These regimes operate at different spatial scales and require different supervision, outputs, and evaluation. A more useful cryoET benchmark should cover both.

\paragraph{Task 1: Compartment Segmentation.} 

This task requires voxel-wise labeling of cellular structures in 3D cryoET volumes, including membrane-bound compartments, filament systems, and other organelles (Fig.~\ref{fig:overview} B(1--2)). This requires models to recover spatially continuous structures across multiple scales, often when boundaries between adjacent structures are weakly defined or obscured by noise.

\paragraph{Task 2: Macromolecular Localization.}

This task focuses on detecting target macromolecules in tomograms and predicting their 3D coordinates, with optional molecular identity labels when multiple species are present (Fig.~\ref{fig:overview} B(3-4)). Localization is a key step in subtomogram averaging pipelines, where detected particles are aligned and averaged to recover higher-resolution structures~\cite{burt2024image,tegunov2019real}.


CryoET segmentation and localization share fundamental challenges arising from the imaging process. Tomograms are noisy, anisotropic, and affected by missing-wedge artifacts, while biological structures vary across samples and imaging conditions. These factors make structural boundaries and discrete object instances difficult to resolve, and lead to strong sensitivity to dataset characteristics and reconstruction artifacts. A more detailed discussion of cryoET data properties and imaging artifacts is provided in Appendix~\ref{imaging_artifacts}.

\begin{figure}
\centering
\includegraphics[width=\linewidth]{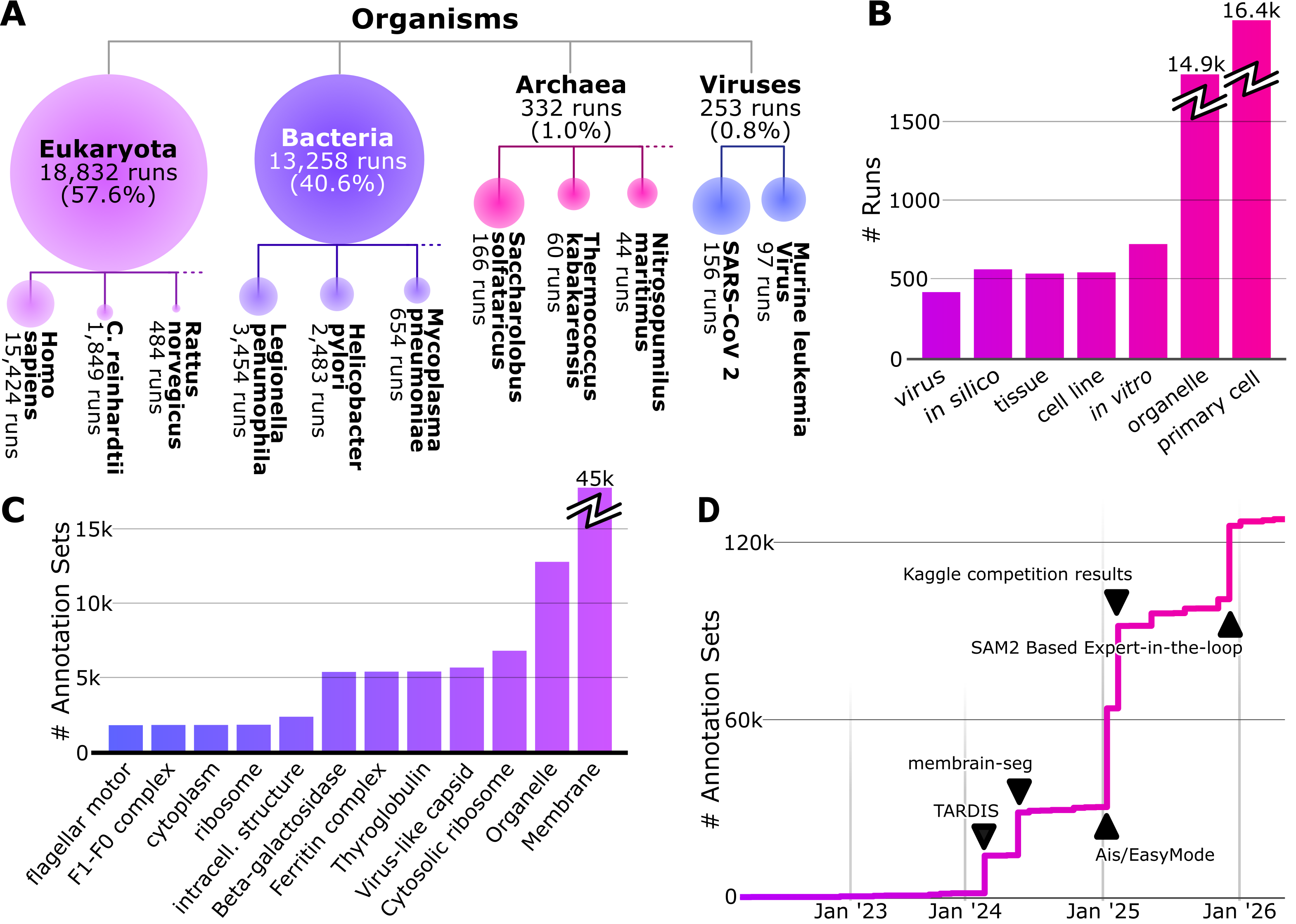}
\caption{
\textbf{Statistical overview of content currently available in the CryoET Data Portal.}
\textbf{(A)} Distribution of available datasets across the tree of life.
\textbf{(B)} Number of runs, where each run corresponds to an individual tomography experiment or replicate, across major biological sample types.
\textbf{(C)} Number of annotation sets available for major annotated targets.
\textbf{(D)} Growth in the total number of annotation sets over time. Major contribution sources are labeled: TARDIS~\cite{kiewisz2022membrane}, membrane-seg~\cite{lamm2024membrain}, Ais/EasyMode~\cite{last2024streamlining}, Kaggle competition results~\cite{peck2025kaggle}, and SAM2-based expert-in-the-loop annotations~\cite{saber2026github}.
}
\label{fig:numbers}
\end{figure}

\section{Datasets}
\label{sec:datasets}

POPSICLE is built from four annotated datasets in the CryoET Data Portal that span the main design axes of cryoET benchmarking: eukaryotic and prokaryotic systems (Fig.~\ref{fig:numbers}A), cellular- and molecular-scale tasks (Fig.~\ref{fig:overview}B, D), and controlled and fully \textit{in situ} imaging settings. Given this large data diversity, the present release covers only a small fraction of the species and molecular targets currently hosted on the portal (Fig.~\ref{fig:numbers}B,C) and is growing as portal annotations expand (Fig.~\ref{fig:numbers}D).

The yeast and bacterial datasets both support semantic segmentation (Task 1, Section~\ref{sec:tasks}) on cellular tomograms, providing annotations for six and five compartment classes, respectively. They offer cellular-scale supervision across biologically and experimentally distinct imaging regimes. The bacterial dataset comprises 68 training tomograms, while the yeast dataset contains 15. This difference in scale creates two distinct learning settings: a relatively well-sampled setting for bacterial segmentation and a sparse, high-variance regime for yeast.

The Phantom and MotorBench datasets support macromolecular localization (Task 2, Section~\ref{sec:tasks}) and cover complementary scenarios. Both were first released through public Kaggle challenges, and the CryoET Data Portal hosts the ground truth labels with top community submissions~\cite{peck2025kaggle,motorbench2025biorxiv}. POPSICLE therefore inherits not only the datasets themselves, but also strong community reference points derived from a broader algorithmic search than any single benchmark study could provide. Per-dataset details are provided in Appendix~\ref{sec:dataset_overview}.

In the Phantom benchmark, target particles span a wide range of sizes and detection difficulty, creating a challenging setting for localization across classes. It contains 492 experimentally acquired tomograms and approximately 60{,}000 expert-curated labels across six particle classes~\cite{peck2025kaggle}. For the challenge, these tomograms were split into 7 training, 121 public test, and 364 private test tomograms; POPSICLE adopts the same split to preserve direct comparability with the challenge setting and its associated community submissions~\cite{peck2025kaggle}. In contrast, MotorBench is a single-class localization objective in intact cellular data, targeting bacterial flagellar motors in whole-cell tomograms~\cite{motorbench2025biorxiv}.

All datasets and annotations used in POPSICLE are accessible through the CryoET Data Portal under the CC0 1.0 license. Dataset sources, identifiers, and provenance are documented in Appendix~\ref{sec:dataset_overview}, the accompanying Croissant metadata files, and the Hugging Face dataset card.

\section{Evaluation Procedures}
\label{sec:eval}

We evaluate a representative set of architectures spanning general-purpose volumetric segmentation models, transformer-based models, and cryoET-specific methods. Our goal is not to optimize each model exhaustively for a single dataset, but to compare strong and widely used model families under a common benchmark setting.

\subsection{Standard Baseline Models}

We evaluate four standard volumetric architectures for both dense segmentation and sparse localization tasks. Our models are 3D nnU-Net, nnU-Net ResEnc, MedNeXt, and SwinUNETR. nnU-Net is a self-configuring framework for biomedical image segmentation and serves as a strong reference model in volumetric imaging~\cite{isensee2021nnu}. nnU-Net ResEnc extends this framework with a residual encoder and deeper feature hierarchy~\cite{isensee2024nnunet}. MedNeXt is a ConvNeXt-inspired convolutional architecture tailored for medical image segmentation~\cite{roy2023mednext}. SwinUNETR combines a hierarchical Swin Transformer encoder with a U-Net-style decoder~\cite{swinunetr}.  Additionally, to incorporate a cryoET-specific design, we also evaluate Octopi, a model family developed for segmentation and sparse object prediction in noisy 3D tomograms. All models are trained using a unified training protocol as described in Appendix~\ref{sec:implementation}.

For benchmarking, we report additional community reference results. On MotorBench, we include top-ranked challenge submissions. For Phantom, we report our trained baselines alongside published in-field methods (DeepFinder~\citep{moebel2021deep} and DeepETPicker~\citep{guoule2024deepetpicker}), as well as the top Kaggle submissions from the original challenge, following the published evaluation protocols~\cite{peck2025kaggle,motorbench2025biorxiv}.

\subsection{Evaluation Metrics}
\label{sec:eval_metric}

We evaluate segmentation performance using the voxel-level Dice score, a standard metric in biomedical imaging~\cite{isensee2021nnu,muller2022towards}. Dice is computed per class and averaged across tomograms.

Localization is evaluated using recall-weighted $F_{\beta}$ scores, following the original challenge protocols for each dataset. The $F_{\beta}$ metric provides a tunable trade-off between precision and recall, allowing the evaluation to adapt to varying levels of annotation completeness and uncertainty. In cryoET, where annotations may be incomplete or ambiguous, this flexibility is important for supporting discovery of previously unlabeled or difficult-to-identify targets.

For the Phantom dataset, we use $F_{4}$ to place strong emphasis on recall. This reflects lower confidence and potential incompleteness in annotations for challenging particle classes, where missing true positives is a greater concern than over-predicting candidates. In contrast, the MotorBench dataset uses $F_{2}$, as the organizers were able to identify and label nearly all the flagellar motors in the volumes, making false positives more meaningful errors~\cite{motorbench2025biorxiv}. Full details of metric computation and evaluation protocols are provided in Appendix~\ref{sec:app_eval}.

\subsection{Reported Results}

We report per-class and aggregate performance separately for segmentation and localization rather than forcing all tasks into a single scalar score. For segmentation, this reveals how models behave on dominant structures versus small or spatially sparse compartments. For localization, it distinguishes relatively easy targets from difficult ones and makes explicit the effect of class imbalance and target morphology.

In addition to our trained baselines, we report challenge-derived community reference points for Phantom and MotorBench. These submissions provide useful upper-bound comparisons because they reflect a broader algorithmic search, including ensembling and extensive inference tuning, than is practical to reproduce within a single unified benchmark study~\cite{peck2025kaggle,motorbench2025biorxiv}. We treat these results as external reference points rather than directly comparable controlled baselines.

Overall, the evaluation is designed to answer two questions. First, how do commonly used model families compare when trained under a shared benchmark protocol? Second, how strongly do model rankings depend on task type, dataset regime, and annotation format? POPSICLE is intended to make both questions measurable within a single reproducible evaluation framework.

\section{Benchmark Results}
\label{results}



For the bacterial dataset, mean Dice scores and their standard deviations are similar across models for most compartments, with the exception of inclusion (Table~\ref{tab:bacterial}). This suggests comparable performance across architectures. Differences are driven by class properties: flagella and inclusions are the most challenging targets, as they occupy less volume and are less frequently present than dominant structures such as the cytoplasm and membrane. 

\begin{table}[ht]
  \centering
  \caption{\textbf{Bacterial segmentation performance.} Per-class Dice scores reported as mean$\pm$standard deviation across held-out test tomograms for each bacterial compartment. IMS: intermembrane space.}
  \label{tab:bacterial}
  \small
  \setlength{\tabcolsep}{10pt}
  \begin{tabular}{lccccc}
    \toprule
    Model & Cytoplasm & Flagella & Membrane & Inclusion & IMS\\
    \midrule
    nnU-Net          & 0.93$\pm$0.04 & 0.52$\pm$0.35 & \textbf{0.86$\pm$0.08} & \textbf{0.74$\pm$0.37} & 0.73$\pm$0.12 \\
    nnU-Net-ResEnc  & \textbf{0.94$\pm$0.04} & \textbf{0.62$\pm$0.29} & 0.86$\pm$0.07 & 0.48$\pm$0.44 & \textbf{0.75$\pm$0.10}  \\
    MedNeXt         & 0.91$\pm$0.09 & 0.61$\pm$0.33 & 0.85$\pm$0.08 & 0.68$\pm$0.36 & 0.74$\pm$0.11  \\
    Octopi            & 0.91$\pm$0.04 & 0.41$\pm$0.33 & 0.80$\pm$0.07 & 0.46$\pm$0.39 & 0.68$\pm$0.14  \\
    SwinUNETR        & 0.91$\pm$0.07 & 0.58$\pm$0.21 & 0.81$\pm$0.07 & 0.72$\pm$0.20 & 0.69$\pm$0.15  \\
    \bottomrule
  \end{tabular}%
\end{table}


In contrast, yeast segmentation is substantially more variable across models and compartments (Table~\ref{tab:yeast}). Large structures such as cytoplasm are segmented reliably, but performance drops for smaller or less prevalent compartments. This suggests that the main challenge is not model capacity alone, but the interaction between object scale, label availability, and imaging artifacts. In particular, the yeast data are affected by a stronger missing wedge and increased anisotropy, making fine or sparse structures harder to recover.

\begin{table}[ht]
  \centering
  \caption{\textbf{Yeast segmentation performance.} Per-class Dice scores reported as mean$\pm$standard deviation across held-out test tomograms for each compartment.}
  \label{tab:yeast}
  \small
  \setlength{\tabcolsep}{4.5pt}
  \begin{tabular}{lcccccc}
    \toprule
    Model &  Cytoplasm & Nucleus & Nuclear Env. & Vesicle & Endosome & Mitochondria \\
    \midrule
    nnU-Net           & 0.89$\pm$0.01 & 0.07$\pm$0.01 & 0.48$\pm$0.02 & 0.02$\pm$0.01 & \textbf{0.53$\pm$0.01} & \textbf{0.49$\pm$0.01} \\
    nnU-Net-ResEnc  & 0.93$\pm$0.04   & 0.24$\pm$0.27   & 0.32$\pm$0.34   & 0.69$\pm$0.21   & 0.47$\pm$0.32   & 0.46$\pm$0.46   \\
    MedNeXt         & \textbf{0.93$\pm$0.02}   & \textbf{0.35$\pm$0.32}   & 0.35$\pm$0.36   & 0.66$\pm$0.11   & 0.43$\pm$0.27   & 0.46$\pm$0.46  \\
    Octopi          & 0.91$\pm$0.04     & 0.24$\pm$0.28     & \textbf{0.50$\pm$0.36}       & \textbf{0.72$\pm$0.13}       & 0.41$\pm$0.31       & 0.44$\pm$0.44 \\
    SwinUNETR  & 0.89$\pm$0.04     & 0.21$\pm$0.21      & 0.32$\pm$0.15       & 0.25$\pm$0.16        & 0.34$\pm$0.19       & 0.39$\pm$0.18 \\
    \bottomrule
  \end{tabular}%
\end{table}



For the localization benchmarks, we report Kaggle competition results alongside standard baseline models. On the MotorBench challenge (Table~\ref{tab:motorbench}), the second-place submission used an nnU-Net-ResEnc architecture, indicating that a standard segmentation model can remain highly competitive for this single-class localization task. In contrast, on the Phantom dataset (Table~\ref{tab:phantom}), the strongest challenge submissions significantly outperform the standard baselines. These methods incorporated extensive task-specific optimization, including substantial data augmentation, model ensembling, and post-processing strategies; see Appendix~\ref{sec:phantom_architectures} for details.

\begin{table}[ht]
    \centering
    \caption{\textbf{MotorBench single-class localization challenge.} Precision,
    recall, and $F_2$ on the private test set. We report top submissions from the BYU Locating Bacterial Flagellar Motors Challenge~\citep{motorbench2025biorxiv}, together with the primary model family used by
  each team.}
    \label{tab:motorbench}
    \small
    \begin{tabular}{l l ccc}
    \toprule
    Rank & Model family & Precision & Recall & \textbf{$F_2$} \\
    \midrule
    1st place        & 3D U-Net               & \textbf{0.784} & \textbf{0.903} & \textbf{0.877} \\ 
    2nd place       & nnU-Net-ResEnc   & \textbf{0.784} & \textbf{0.903} & \textbf{0.877} \\ 
    3rd place       & Hybrid 3D/2D           & 0.770 & 0.894 & 0.866 \\ 
    4th place        & Hybrid ResNet-18 detection  & 0.767 & 0.894 & 0.865 \\ 
    5th place & 2.5D YOLOv8           & 0.780 & 0.880 & 0.858 \\ 
    \bottomrule
  \end{tabular}
\end{table} 

\begin{table}[ht]
  \centering
  \caption{\textbf{Multi-class localization challenge on the Phantom dataset.} Per-class $F_4$ and weighted aggregate $\overline{F}_4$ on the Phantom private test set (Dataset ID: DS-10446). We report our baselines alongside published in-field methods and the top-ranked submissions from the CZII Kaggle Challenge~\citep{peck2025kaggle}. VLP: virus-like particle.}
  \label{tab:phantom}
  \small
  \begin{tabular}{l cccccc c}
    \toprule
    & Apo. & Rib. & VLP & $\beta$-gal & Thyro. & $\beta$-amyl.
    & $\overline{F}_4$ \\
    \midrule
    \multicolumn{8}{l}{\textit{Our baselines}} \\ 
    nnU-Net         & 0.706 & 0.803 & 0.940 & 0.561 & 0.482 & 0.285 & 0.648 \\
    nnU-Net ResEnc   & 0.604 & 0.676 & 0.929 & 0.212 & 0.266 & 0.111 & 0.452 \\
    MedNeXt          & 0.614 & 0.735 & 0.940 & 0.0 & 0.240 & 0.0 & 0.396 \\
    SwinUNETR         &0.663 &0.593 &0.754 &0.376 &0.337 &0.176 &0.491 \\
    Octopi            & 0.896 & 0.879 & 0.922 & 0.590 & 0.571 & 0.459 & 0.763 \\
    \addlinespace[5pt]
    \midrule
    \multicolumn{8}{l}{\textit{Published cryoET-specific methods}} \\
    DeepFinder~\citep{moebel2021deep}         & 0.434 & 0.790 & 0.880 & 0.434 & 0.498 & 0.0 & 0.567 \\
    DeepETPicker~\citep{guoule2024deepetpicker}        & 0.731 & 0.851 & 0.941 & 0.533 & 0.571 & 0.0 & 0.676 \\
    \addlinespace[5pt]
    \midrule
    \multicolumn{8}{l}{\textit{CZII Kaggle Challenge}} \\
    1st place  & \textbf{0.932} & \textbf{0.912} & 0.952 & \textbf{0.688} & 0.671 & 0.205 & \textbf{0.788} \\
    2nd place            & 0.931 & 0.911 & 0.943 & 0.677 & 0.674 & 0.0 & 0.784 \\
    3rd place            & \textbf{0.932} & 0.908 & 0.952 & 0.669 & \textbf{0.678} & 0.0 & 0.784 \\
    5th place            & 0.928 & \textbf{0.912} & 0.943 & 0.681 & 0.666 & 0.0 & 0.783 \\
    10th place           & 0.930 & 0.893 & \textbf{0.955} & 0.662 & 0.653 & \textbf{0.465} & 0.773 \\
    \bottomrule
  \end{tabular}
\end{table}

A key finding across experiments is that performance does not transfer reliably from segmentation to localization. Architectures optimized for dense voxel-wise prediction struggle to separate nearby instances and produce accurate detections, particularly under extreme class imbalance. For example in the Phantom Kaggle challenge, nnU-Net achieves an $\overline{F}_4$ score of 0.648 , corresponding to approximately 580th place out of 931. This illustrates a substantial drop in performance when transitioning from segmentation to particle localization.

We find performance is also class-dependent: large, high-contrast regions are segmented reliably, whereas thin, low-contrast, or spatially sparse structures remain challenging for all methods (e.g., Nucleus and $\beta$-amylase). This pattern indicates that current models struggle with resolving fine-scale features under anisotropic resolution and limited signal-to-noise conditions, a core challenge in cryoET.


Taken together, we see that model rankings are highly task- and dataset-dependent. Architectures that are competitive in one setting often underperform in others, and no single approach consistently leads across segmentation and localization. This contrasts with trends in adjacent domains such as medical imaging, where a small set of closely related architectures consistently dominate. For example, in the Touchstone benchmark MedNeXt, nnU-Net, and related U-Net–style variants consistently rank as top performers across organs and datasets, indicating stable architectural hierarchy~\cite{touchstone_benchmark}. 


\section{Discussion \& Conclusion}
\label{sec:conclusion}

POPSICLE introduces a unified benchmark for cryoET that evaluates dense segmentation and sparse macromolecular localization across multiple biological and experimental regimes. Our results show that performance depends strongly on both task and dataset, and that no single architecture is consistently strong across all settings. In particular, models that perform well on voxel-wise segmentation often degrade on localization, indicating that current approaches do not transfer reliably between dense and sparse prediction regimes.

More broadly, the benchmark shows that evaluation on a single dataset or task can give a misleading picture of model capability. Performance differences across POPSICLE datasets reflect not only biological variation, but also differences in data scale, object prevalence, and structural complexity. Robust evaluation therefore requires testing across multiple tasks to expose complementary strengths and failure modes.

This gap highlights an important direction for future cryoET method development. Segmentation and localization place different demands on a model: segmentation requires recovering spatially continuous structures under dense supervision, whereas localization requires identifying sparse discrete targets under extreme class imbalance. In practice, however, cryoET analysis often requires both capabilities within the same tomogram, since cellular context and molecular localization are complementary parts of biological interpretation. A unified model or training strategy that can operate across these regimes would reduce the need for task-specific pipelines, make better use of shared structural information, and provide a more scalable foundation as annotations in the CryoET Data Portal continue to expand. POPSICLE provides a benchmark setting for measuring progress toward this goal, even though the present results show that current models do not yet achieve it.

POPSICLE is designed as a living benchmark built on the CryoET Data Portal. As new datasets and annotations are added to the portal, the benchmark will expand in biological scope, imaging diversity, and task coverage. This is important because realistic cryoET evaluation will increasingly require harder settings, including thin filamentous structures, rare targets, and context-dependent molecular complexes. The integration of segmentation, localization, and metadata also creates a foundation for more general cryoET models, including promptable or text-guided systems such as VoxTell~\cite{rokuss2025voxtell}.

The current release has some limitations. It covers only a subset of the biological diversity and annotation types relevant to cryoET, with emphasis on semantic segmentation and point localization rather than instance segmentation, oriented targets, or filament tracing. Like other cryoET resources built from expert annotation, it is also affected by label noise arising from annotation uncertainty, incomplete ground truth, and differences in annotation conventions across datasets. Our experiments compare representative model families under a shared protocol, but do not attempt exhaustive per-model optimization. In contrast, the challenge-derived community submissions reflect extensive task-specific tuning, ensembling, and post-processing, and should therefore be interpreted as reference points rather than controlled baselines. In addition, because POPSICLE is built on an evolving data resource, its current scope should be understood as a snapshot rather than a fixed endpoint.


We do not identify reasonably foreseeable negative societal impacts from the release or use of POPSICLE beyond ordinary scientific risks of misinterpreting benchmark results, which are addressed through documentation of dataset scope, limitations, and intended use in the paper and accompanying dataset materials. Overall, POPSICLE is intended to make cryoET evaluation more reproducible, broader in scope, and easier to extend as the underlying data resource grows. We hope it will support more consistent comparison of methods and the development of models that generalize across the multi-scale and heterogeneous structure of cryoET data.

\section{Acknowledgements}

We sincerely thank David Agard for insightful discussions, the HPC team at Biohub for computational support, and Rachel Webb for writing assistance. We thank Pallavi Khedle, Daniel Ji, Josh Hutchings, Rahel Woldeyes, David Dong, and Mykhailo Kopylov for their assistance with tomogram annotation. This work was supported by Biohub and its donors, Priscilla Chan and Mark Zuckerberg. 

\bibliographystyle{unsrtnat}
\bibliography{bibliography}

\newpage

\appendix

\section*{Appendix}

This appendix provides detailed insights into the POPSICLE benchmark and is organized as follows:

\begin{itemize}

    \item Appendix~\ref{sec:dataset_overview}: Dataset overview and access details via the CryoET Data Portal.

    \item Appendix~\ref{architectures}: Benchmark model architectures evaluated in this work, including their design and adaptation to CryoET tasks.

    \item Appendix~\ref{sec:data_structure}: Data structure, including formats for tomograms, annotations, and dataset organization.

    \item Appendix~\ref{imaging_artifacts}: Key characteristics of CryoET data, including imaging artifacts and domain-specific challenges that affect model performance.

    \item Appendix~\ref{sec:tools}: Supporting tools and infrastructure for data access, preprocessing, and benchmarking.

\end{itemize}

\section{Dataset Overview}
\label{sec:dataset_overview}

We provide a summary of the datasets included in POPSICLE, covering both segmentation and localization tasks across multiple biological and imaging regimes. Segmentation datasets in POPSICLE consist of cryo-electron tomograms paired with multi-label voxel-wise annotations defined on a shared three-dimensional grid, where each tomogram is associated with a segmentation mask of identical shape for direct voxel-level supervision. Localization datasets consist of tomograms paired with point annotations specifying the 3D coordinates of target molecular complexes. Table~\ref{tab:datasets} summarizes the key properties of each dataset, including task type, organism, annotation modality, and dataset identifiers in the CryoET Data Portal.

\begin{table}[ht]
\centering
\caption{\textbf{Overview of POPSICLE benchmark datasets.} Summary of datasets, tasks, annotation types, and scale. The Phantom dataset has an additional 121 annotated tomograms in the test/validation split, yielding a total of 2,993 tomograms.}
\label{tab:datasets}
\small
\begin{tabular}{p{1.5cm} p{1.8cm} p{2.2cm} p{2.5cm} c c c}
\hline
\textbf{Dataset} & \textbf{Task} & \textbf{Organism} & \textbf{Annotation Type} & \textbf{\# Train} & \textbf{\# Test} & \textbf{\# Classes} \\
\hline
Yeast & Segmentation & Yeast & Dense voxel-wise & 15  & 4 & 6 \\
Bacterial & Segmentation & Prokaryote & Dense voxel-wise & 68 & 12  & 5 \\
Phantom & Localization & Lysate / synthetic & Points & 7 & 364 & 6 \\
MotorBench & Localization & Prokaryote & Points & 1,559  & 843 & 1 \\
\hline
\end{tabular}
\end{table}

To facilitate reproducibility, we distinguish between dataset-level identifiers and higher-level depositions used for larger collections. Table~\ref{tab:data_access} summarizes the CryoET Data Portal access identifiers for all datasets used in POPSICLE. 

\begin{table}[h]
\caption{\textbf{CryoET Data Portal access details for POPSICLE datasets.} DatasetIDs (DS-XX) and larger collections are referenced via deposition IDs (CZCDP-XX).}
\centering
\small
\begin{tabular}{l l l l}
\toprule
\textbf{Dataset} & \textbf{Access Type} & \textbf{Identifier(s)} & \textbf{Notes} \\
\midrule
Yeast & Dataset IDs & DS-10000, DS-10001, CZCDP-10351 & Data split in croissant \\
Bacterial & Deposition & CZCDP-10350 & Data split in croissant \\
Phantom & Dataset IDs & DS-10440, DS-10445, DS-10446 & Train / Validation / Test \\
MotorBench & Deposition & CZCDP-10332 / CZCDP-10347 & Train / Test \\
\bottomrule
\end{tabular}
\label{tab:data_access}
\end{table}

We next describe each dataset in detail, focusing on annotation structure and dataset-specific characteristics.

\subsection{\textit{Schizosaccharomyces pombe} Yeast Segmentation}
The yeast segmentation dataset is derived from previously published cryoET annotations of \textit{S. pombe}~\cite{deepict} and is available through the CryoET Data Portal under deposition ID CZCDP-10351 (tomograms available under dataset IDs DS-10000 and DS-10001).

The annotations cover six cellular compartments: cytoplasm, nucleus, nuclear envelope, vesicle, endomembrane, and mitochondrion. These classes are unevenly distributed across the dataset, with several compartments appearing in only a subset of tomograms. In particular, small compartments such as vesicles and nuclei occupy limited spatial volume and are absent in many samples, resulting in strong class imbalance. 

\subsection{Bacterial Segmentation}
\label{sec:app_bacterial}

The bacterial segmentation dataset spans multiple prokaryotic species and is available through the CryoET Data Portal under deposition CZCDP-10350, comprising 80 annotated tomograms.

The cellular compartments and flagella in bacterial tomograms were initially annotated using napari-nnInteractive~\cite{Isensee2025-iq}. Membrane segmentations were obtained from the union of CryoET Data Portal depositions CZCDP-10301 and CZCDP-10303. Individual compartment, flagellar and membrane annotations were then combined using a curation pipeline created using copick-MCP with Claude Opus 4.6 (Anthropic, San Francisco, CA), as described previously~\cite{copick}. As a final step, the combined set of segmentations was proofread manually by an expert annotator for each tomogram. 

The annotations provide five structural classes: cytoplasm, membrane, intermembrane space, flagellum, and inclusion. Core structures such as cytoplasm and membranes are present in all tomograms, while flagella and inclusions occur less frequently, introducing moderate class imbalance. The dataset spans multiple bacterial genera and is partitioned into 68 training and 12 test tomograms, split to normalize species and class coverage.

Relative to the yeast dataset, this dataset represents a well-sampled regime with more uniform class coverage and larger training scale. As a result, it enables more stable model training and evaluation while still presenting challenges for fine-scale structures such as the intermembrane space due to low contrast and anisotropic resolution. Within POPSICLE, it serves as a baseline setting for dense segmentation under favorable data conditions.

\subsection{The Phantom Object Localization}

The Phantom dataset is a multi-class macromolecular localization benchmark designed to approximate \textit{in situ} particle detection while remaining scalable for annotation. It consists of cryo-electron tomograms paired with point annotations indicating the 3D coordinates of target molecular complexes across six classes. The dataset is derived from an experimentally acquired lysate sample enriched for lysosomal components, with additional purified targets introduced to control object diversity and class balance~\cite{peck2025kaggle}. Collectively, these six molecular targets spanned over an order of magnitude in molecular weight (268-4300 kilodaltons) and were characterized by different shapes to encourage annotation algorithms that generalize to diverse molecular species.

The dataset contains 492 tomograms split into 7 training, 121 public test, and 364 private test volumes following the original challenge protocol. Corresponding dataset IDs in the CryoET Data Portal are 10440 (train), 10445 (public test) and 10446 (private test). 

The small training set reflects realistic annotation constraints in cryoET, while the large held-out test set enables robust evaluation. Targets span a wide range of molecular sizes and shapes and are embedded within a crowded and heterogeneous background containing endogenous structures that act as natural decoys. As a result, Phantom represents a challenging regime for multi-class localization under limited supervision, requiring models to generalize across object scale, morphology, and spatial context. 

\subsection{MotorBench Flagellar Motor Localization}

MotorBench is a single-class localization dataset targeting bacterial flagellar motors in whole-cell cryo-electron tomograms. Annotations are provided as 3D coordinates corresponding to flagella motors, enabling evaluation of detection performance in realistic \textit{in situ} cellular environments.

The dataset includes 844 held-out test tomograms from \textit{Vibrio cholerae}, of which 327 contain at least one annotated motor and 517 contain none, creating a highly imbalanced detection setting~\cite{czdp10347_motorbench_test}. The training data are drawn from an expanded corpus assembled during and after the associated Kaggle competition, comprising over 2,000 tomograms across multiple bacterial and archaeal species~\cite{czdp10332_motorbench_train}. These annotations are accessible via CryoET Data Portal depositions CZCDP-10332 (training) and CZCDP-10347 (test), providing flagellar motor locations across 91 and 5 CryoET Data Portal datasets, respectively.

Compared to Phantom, MotorBench focuses on a simpler binary detection task but introduces increased biological variability and more complex cellular context. These data combine three sources: the original BYU competition release (\cite{byu2025flagellar_motors_kaggle}), a large external dataset released by the first-place team (\cite{artley2025flagellar_motors_dataset,artley2025byu_competition_solution}), and a corrected and expanded dataset from the MIC-DKFZ team (\cite{micdkfz2025flagellar_motors_solution}). This makes it a useful benchmark for evaluating robustness to distribution shift and performance in realistic experimental conditions. 

\section{Benchmark Model Architectures}
\label{architectures}

We evaluate a set of representative architectures spanning convolutional, transformer-based, and cryoET-specific designs. These models were selected to cover a range of inductive biases and modeling strategies commonly used in volumetric biomedical imaging and cryoET analysis.

\subsection{CNN Architectures}

\textbf{nnU-Net.} nnU-Net~\cite{isensee2021nnu} is a self-configuring framework for biomedical image segmentation that automatically adapts preprocessing, network architecture, and training parameters to a given dataset. We use the 3D full-resolution configuration, which serves as a strong and widely adopted baseline for volumetric segmentation.

\textbf{nnU-Net ResEnc.} nnU-Net ResEnc~\cite{isensee2024nnunet} extends nnU-Net with residual encoder blocks and a deeper hierarchical feature representation. This design increases model capacity and improves gradient flow, enabling better performance on complex volumetric structures.

\textbf{MedNeXt.} MedNeXt~\cite{roy2023mednext} is a convolutional architecture inspired by ConvNeXt and adapted for 3D medical imaging. It employs large kernel convolutions and modern design choices such as inverted bottlenecks and layer normalization to improve performance on volumetric tasks.

\subsection{Transformer Architectures}

\textbf{SwinUNETR.} SwinUNETR~\cite{swinunetr} is a hybrid transformer–CNN architecture that combines a Swin Transformer encoder with a U-Net-style decoder. It leverages hierarchical self-attention to model long-range dependencies while retaining spatial resolution through skip connections.

\subsection{CryoET-Specific Architectures}

\textbf{Octopi.} Octopi~\cite{octopi} is a cryoET-specific framework designed for sparse object localization in highly noisy 3D tomograms. Rather than relying on a fixed architecture, Octopi adopts a self-configuring paradigm that automatically adapts network design, training parameters (i.e., loss function), and inference strategies to the target dataset. This is achieved through exploration of the architectural search space using Bayesian optimization, enabling the selection of model configurations that are well-matched to the underlying data distribution and task structure. 

In addition to architectural adaptation, Octopi incorporates design elements tailored to cryoET localization, including dense-to-sparse prediction schemes, class imbalance handling, and post-processing pipelines for converting voxel-wise outputs into discrete particle coordinates.

\textbf{DeepFinder.}
DeepFinder~\cite{moebel2021deep} is a convolutional architecture designed for macromolecular localization in cryoET volumes. It operates on dense voxel predictions that are post-processed into discrete particle detections, enabling end-to-end learning for object identification in noisy 3D data.

\textbf{DeepETPicker.} DeepETPicker~\cite{guoule2024deepetpicker} extends earlier convolutional particle-picking approaches such as DeepFinder by incorporating deeper residual architectures and improved training strategies for volumetric detection. In contrast to the relatively shallow CNN design of DeepFinder, DeepETPicker introduces residual connections that enable more effective gradient propagation and richer hierarchical feature learning in 3D volumes. 

\subsection{Phantom Challenge Architectures}
\label{sec:phantom_architectures}

These top-performing solutions are dominated by fully 3D-CNNs with performance driven largely by ensembling and task-specific post-processing. While some hybrid and 2.5D approaches (e.g., using ImageNet-pretrained 2D encoders with slice-wise inference and depth aggregation) were explored in the challenge, the top performers were trained from scratch and did not rely on natural-image pre-training.

\begin{table}[h]
\centering
\caption{Top-ranked teams in the CZII Kaggle Challenge}
\label{tab:phantom_challenge_architectures}
\begin{tabular}{ll}
\toprule
Rank & Model family \\
\midrule
1st  & \makecell[l]{An ensemble of segmentation models (3D U-Nets with 3D CNN encoders) \\ 
and YOLO-style object detection models} \\
\addlinespace[3pt]
2nd  & 3D U-Nets with 2D/3D CNN encoders  \\
\addlinespace[3pt]
3rd  & 3D U-Nets with 3D CNN encoders \\
\addlinespace[3pt]
5th  & 3D U-Nets  \\
\addlinespace[3pt]
10th & 3D U-Nets \\
\bottomrule
\end{tabular}
\end{table}

Despite their strong performance, we do not include these designs as benchmark models. Most solutions consist of highly specialized pipelines combining custom inference heuristics, and dataset-specific tuning (e.g., thresholding, clustering, or class-wise post-processing) that are tightly coupled to the Phantom dataset. These factors make them difficult to standardize, reproduce, and apply consistently across new datasets within a unified benchmark setting. Instead, we treat them as community reference points that reflect the upper bound achievable with extensive task-specific optimization, rather than as directly comparable model families under a shared training protocol.

\subsection{Implementation Details}
\label{sec:implementation}

To ensure a fair and reproducible comparison, all models are trained using their respective reference implementations, preserving the intended design and optimization strategies of each method. Table~\ref{tab:model_summary} summarizes the benchmarked models, including parameter counts, implementation frameworks, and tuning strategies. Specifically, nnU-Net, nnU-Net ResEnc, and MedNeXt are trained using the nnU-Net framework and its self-configuration pipeline; SwinUNETR is implemented using MONAI; and Octopi is trained using its native framework with automated model exploration. Model sizes range from 3.26M to 383.5M parameters, reflecting substantial variation across architectures.

\begin{table}[h]
\caption{Summary of benchmarked models, implementation frameworks, and tuning strategies in this work. All models are 3D and are trained under a unified protocol while preserving their native optimization procedures.}
\centering
\small
\begin{tabular}{lcccc}
\toprule
\textbf{Model} & \textbf{Parameters} & \textbf{Category} & \textbf{Framework} & \textbf{Tuning Strategy} \\
\midrule
nnU-Net & 88.2M & CNN & nnU-Net & Self-configuration \\
nnU-Net ResEnc & 383.5M & CNN & nnU-Net & Self-configuration \\
MedNeXt & 10.5M & CNN & nnU-Net & Self-configuration \\
Octopi & 3.26M & CNN & Custom & Bayesian optimization \\
SwinUNETR & 62.2M & Transformer-CNN & MONAI & Octopi-tuned \\
\bottomrule
\end{tabular}
\label{tab:model_summary}
\end{table}

Training follows the default configuration procedures of each framework where applicable (e.g., nnU-Net self-configuration and Octopi model search), rather than extensive manual hyperparameter tuning. For SwinUNETR, we initialize training using hyperparameters identified through the Octopi framework, providing a strong configuration without task-specific manual optimization.

\subsection{Training Protocols}
\label{sec:training}

Models are trained independently on each dataset. To satisfy GPU memory constraints, training is performed on cropped 3D sub-volumes sampled from full tomograms. For segmentation tasks, supervision is provided as voxel-wise semantic masks. For localization tasks, point annotations are converted into dense training targets compatible with each model family, allowing both segmentation-style and localization-specific architectures to be trained under a unified framework.

We use a consistent training protocol across models whenever supported by the underlying implementation. Data augmentation includes random rotations, flips, and intensity perturbations. For segmentation-based models, optimization combines Dice-style overlap losses with cross-entropy supervision where appropriate. For self-configuring frameworks such as nnU-Net, we retain default configuration and training procedures rather than introducing dataset-specific manual tuning, ensuring a fair and reproducible comparison across architectures.

\subsection{Evaluation Protocols}
\label{sec:app_eval}

\paragraph{Segmentation.}
Segmentation performance is evaluated using the voxel-level Dice score. For each class, Dice is computed between predicted and ground-truth segmentation masks over all voxels in a tomogram. Scores are then averaged across classes and across tomograms to produce the final reported metrics.

\paragraph{Localization.}
For localization tasks, model outputs are converted into discrete 3D coordinates using post-processing (e.g., 3D connected components). Where required, class-specific thresholds are selected on validation data and fixed at test time.

For evaluation, predicted coordinates are matched to ground-truth annotations using a distance-based criterion, where each ground-truth point can be matched to at most one prediction. A prediction at location $\bar{y}$ is matched to a ground-truth point $y$ if $\|y - \bar{y}\|_2 \leq \tau$.

Based on this matching, we define true positives (TP) as matched predictions, false positives (FP) as unmatched predictions, and false negatives (FN) as unmatched ground-truth points, from which precision, recall, and the $F_\beta$ score are computed as follows:

\begin{align*}
    \text{Precision} =& \frac{\text{TP}}{\text{TP} + \text{FP}}, \quad \text{Recall} = \frac{\text{TP}}{\text{TP} + \text{FN}} \\
    F_{\beta} &= (1 + \beta^2)\,\frac{\text{Precision} \cdot \text{Recall}}{\beta^2 \cdot \text{Precision} + \text{Recall}} 
\end{align*}

Following the main text, we use different $\beta$ values to reflect differences in annotation completeness across datasets, with higher $\beta$ emphasizing recall in settings with uncertain or incomplete annotations.

\paragraph{MotorBench.}
The MotorBench benchmark evaluates single-class localization using a $F_{2}$ score, with a fixed distance threshold of $\tau = 100$ nm.

\paragraph{Phantom challenge.}
The Phantom benchmark evaluates multi-class localization with a $F_{4}$ score, following the original challenge protocol. For each particle class $c$, predictions are matched to ground-truth annotations using a distance-based criterion, where the matching threshold is set to half the particle radius (ranging between 6--15\,nm across classes), and aggregated across all tomograms to compute a single $F_4^{(c)}$ score.
This corresponds to a micro-averaged computation within each class. The final aggregate score is then obtained by combining per-class scores using a class-weighted average:
\[
\overline{F}_{4} = \frac{\sum_{c \in \mathcal{C}} w_c \, F_{4}^{(c)}}{\sum_{c \in \mathcal{C}} w_c},
\]
where $w_c$ are class-specific weights. As a result, the overall metric is not a micro-average across all detections, but a weighted macro-average over per-class $F_{4}$ scores.

The weighting emphasizes challenging classes and prevents easier particles with high contrast from dominating the aggregate score. In the official evaluation, thyroglobulin and $\beta$-galactosidase are upweighted, while virus-like particle (VLP), ribosomes, and apoferritin receive lower weight. The $\beta$-amylase class is excluded from the aggregate score due to lower confidence in its reference annotations, although it is included in per-class reporting. We follow this protocol exactly when computing $\overline{F}_{4}$.

\begin{table}[htbp]
\centering
\caption{Class weights used in the Phantom $\overline{F}_{4}$ score.}
\label{tab:phantom_weights}
\begin{tabular}{l c}
\toprule
\textbf{Class} & \textbf{Weight $w_c$} \\
\midrule
Virus-like particle & 1 \\
Ribosome & 1 \\
Apoferritin & 1 \\
Thyroglobulin & 2 \\
$\beta$-galactosidase & 2 \\
$\beta$-amylase & 0  \\
\bottomrule
\end{tabular}
\end{table}

\subsection{Runtime Details}
\label{sec:runtime}

We report training compute and optimization configurations for all benchmarked models in Tables~\ref{tab:compute_config} and~\ref{tab:training_config}. Unless otherwise noted, training is performed using a consistent hardware setup across both segmentation and localization tasks. In particular, all models are trained on NVIDIA A6000 or H100 GPUs with memory configurations ranging from 48GB to 80GB, and identical training settings are used across tasks to ensure fair comparison. This reflects our unified training protocol, where localization is formulated through dense supervision compatible with segmentation architectures (see Appendix~\ref{sec:training}).

\begin{table}[htbp]
\centering
\caption{Training hyperparameters for each architecture.}
\label{tab:training_config}
\small
\setlength{\tabcolsep}{5pt}
\begin{tabular}{lcccccc}
\toprule
\textbf{Architecture} & \textbf{Patch Size} & \textbf{Batch Size} & \textbf{Optimizer} & \textbf{Loss} & \textbf{LR, Scheduler} \\
\midrule
nnU-Net        & $96{\times}160{\times}160$ & 2 & SGD & Dice, CE & 1e-2, PolyLR \\
nnU-Net ResEnc & $128{\times}256{\times}224$ & 2 & SGD & Dice, CE & 1e-2, PolyLR \\
MedNeXt        & $96{\times}160{\times}160$ & 2 & SGD & Dice, CE & 1e-2, PolyLR \\
SwinUNETR      & $96{\times}96{\times}96$ & 16 & AdamW & Focal Tversky & 1e-4, CosineAnnealing \\
Octopi         & $128{\times}128{\times}128$ & 48 & AdamW & Focal & 5e-4, Cosine \\
\bottomrule
\end{tabular}
\end{table}

For community reference results derived from Kaggle challenges, we do not report training time or total compute, as these submissions were produced under heterogeneous and non-standardized training environments. However, we note that Kaggle competitions provide constrained inference environments consisting of two NVIDIA T4 GPUs with a fixed 12 hour time limit. Reported challenge results therefore reflect not only model design but also optimization under these resource constraints.

\begin{table}[htbp]
\centering
\caption{Average training compute and runtime configuration per model. SwinUNETR iterates over all training tomograms for each task resulting in broad training time proportional to dataset size.}
\label{tab:compute_config}
\begin{tabular}{lccccc}
\toprule
\textbf{Architecture} & \textbf{Epochs} & \textbf{Hours/GPU} & \textbf{GPU Type} & \textbf{Avg Memory/GPU} \\
\midrule
nnU-Net        & 1,000 & 17.8 & H100 & 13.5GB \\
nnU-Net ResEnc & 1,000 & 57 & H100 &  22.8GB \\
MedNeXt        & 1,000 & 31.5 & H100 & 24.2GB \\
SwinUNETR      & 300-400 & 8-45 & $4\times$ H100 & 52.3GB \\
Octopi         & 1,000 & 6.92 & A6000 / H100 & 28.5GB \\
\bottomrule
\end{tabular}
\end{table}

Overall, these tables are intended to provide transparency into the computational requirements of our controlled experiments, while distinguishing them from externally sourced challenge submissions that follow different compute regimes.

\section{Data Structure}
\label{sec:data_structure}

CryoET datasets begin as a series of 2D movie stacks acquired by tilting a vitrified biological specimen over a fixed tilt axis within a transmission electron microscope. Each tilt image is recorded as a short movie to allow correction for beam-induced motion \cite{peck2025aretomolive}. These raw movies are first processed through motion correction, generating a single high-quality image per tilt. The resulting tilt series—comprising these corrected 2D projections—are then reconstructed into 3D tomograms using physics-based algorithms (Fig.~\ref{fig:overview}A). The CryoET Data Portal employs AreTomo3~\cite{aretomo3}, a GPU-accelerated reconstruction tool that performs global and local motion correction, tilt-series alignment, contrast transfer function (CTF) correction, and weighted back-projection to produce tomograms at multiple voxel sizes. For improved interpretability, tomograms are optionally denoised in real time using DenoisET~\cite{aretomo3}, a self-supervised denoising model based on Noise2Noise.

To accommodate users with varying levels of technical expertise and computational resources, the CryoET Data Portal provides data at multiple intermediate stages: raw movie stacks, motion-corrected tilt series, reconstructed tomograms (with and without denoising), annotations, and associated metadata. This enables both reproducible benchmarking and flexible integration into user-defined workflows.

In addition, the portal supports downstream geometric preprocessing for masking and sample localization. Using tools like copick (see Appendix \ref{sec:tools}), users can prepare high-quality training data for machine learning based localization, identification and segmentation.

By aligning its preprocessing pipeline with state-of-the-art cryoET reconstruction practices and packaging it into scalable, modular software components, the CryoET Data Portal lowers the entry barrier for large-scale annotation and algorithm development. It provides a full-stack processing ecosystem that spans raw acquisition data to ML-ready 3D volumes.

Accurately identifying the physical boundaries of biological samples within cryoET tomograms is essential for tasks such as masking, training data preparation, and quality control. Since cryoET reconstructions often include large volumes of vacuum surrounding the sample, predicting the true extent of the specimen can significantly reduce data size and improve downstream model performance. Using copick integrated with ChimeraX for interactive point-based annotation, users annotate the top and bottom surfaces of the sample and fit smooth boundary meshes to outline the valid reconstruction volumes. This yields dense segmentations and training targets for downstream tools, which enables automated and reproducible sample localization while remaining modular.

{\bfseries Metadata structure: }
The CryoET Data Portal is organized into a structured metadata hierarchy. Data is grouped into datasets (defined by shared sample and preparation conditions), which contain multiple runs (individual imaging experiments), each of which may include several tomograms and annotations. Rich metadata is collected at every level—including acquisition parameters, sample context, processing steps, and annotation methods—and stored in standardized JSON files compliant with a public LinkML schema. This metadata enables users to search, filter, and download data via the portal interface or API, and supports downstream integration with tools like Neuroglancer and Napari. By enforcing metadata consistency and completeness, the portal facilitates robust data curation, cross-dataset comparisons, and machine learning model training on biologically meaningful subsets.

{\bfseries Runs: }
A run represents all data and annotations from imaging a single location in a sample and belongs to a dataset that may include multiple runs. Each run is linked to its tomograms—often at multiple voxel spacings—and is identified by a unique Run ID. Metadata for each run is accessible via its Run Overview Page on the Portal.

{\bfseries Tilt series: }
Each tilt series typically includes 30–50 tilts per run. These frames are corrected for beam-induced motion and summed to generate a single image per tilt, forming the tilt series. Associated metadata—such as acquisition parameters, gain references for detector calibration, and tilt angles are provided alongside the tilt series in standardized formats (MRC~\cite{cheng2015mrc2014} and OME-Zarr~\cite{omezarr}). Each tilt series has a unique ID and an author-assigned quality score (1–5), reflecting its alignment accuracy and usability for downstream analysis such as tomogram reconstruction.

{\bfseries Tomograms: }
Tomograms in the CryoET Data Portal are 3D reconstructions generated from aligned tilt series, often corrected for electron optics aberrations. Each tomogram has a unique ID and is linked to detailed metadata including voxel spacing, reconstruction method, and any post-processing steps such as denoising. Alignment metadata (stored as JSON files) includes affine transformations and alignment type. Tomograms are listed on Run Overview pages, where users can visualize them (with annotations) via Neuroglancer or download them in MRC or OME-Zarr format, along with programmatic access options.

{\bfseries Annotations: }
Annotations in the CryoET Data Portal identify macromolecular structures within tomograms and are organized by object type (e.g., ribosome, membrane) and shape type—segmentation, instance segmentation, point, or oriented point. Each annotation has a unique ID and includes metadata about the annotation method (manual, automated, or hybrid), software used, and optional confidence metrics such as precision and recall when ground truth is available. Annotations are visualized alongside tomograms in Neuroglancer, with default layers curated to avoid occlusion and emphasize ground truth when present. Users can download annotations in browser-friendly JSON formats or as volumetric masks (MRC/OME-Zarr) using the Portal interface, API, or AWS CLI. Annotations marked as ground truth are flagged for use in machine learning training and validation.

\section{CryoET Data Characteristics and Imaging Artifacts}
\label{imaging_artifacts}

Cryo-electron tomography poses unique challenges for machine learning due to its characteristic data properties and imaging artifacts. Unlike natural images or even other biomedical imaging modalities, cryoET data is acquired by collecting a tilt series of 2D projections under extremely low electron dose, followed by tomographic reconstruction. This results in noisy, anisotropic 3D volumes with several key constraints:

{\bfseries Low Signal-to-Noise Ratio (SNR): }
To minimize radiation damage to vitrified biological samples, imaging is performed with extremely low electron doses. This results in SNRs orders of magnitude lower than those seen in standard computer vision datasets, with many features visually indiscernible.

{\bfseries Missing Wedge Artifact: }
Due to physical limitations on the tilt range during acquisition (typically $-60^\circ$ to $+60^\circ$), tomogram reconstructions suffer from a cone of missing information in Fourier space. This induces anisotropic resolution, elongation of features, and systematic bias in 3D structure.

{\bfseries Sample Heterogeneity: }
Biological variability is intrinsic to cryoET datasets, with differences arising from cell type, experimental conditions, and molecular states. Moreover, macromolecular complexes often appear in diverse orientations and conformations, adding complexity to downstream tasks like segmentation or classification.

{\bfseries Spatially Varying Contrast and Thickness: }
Variability in sample thickness and orientation across the field of view leads to contrast variation and uneven image quality, further complicating global normalization and data augmentation strategies.

{\bfseries Tomogram-Specific Reconstruction Artifacts: }
Reconstruction methods (e.g., weighted back projection, SART) introduce distinct artifacts depending on parameter choices and preprocessing pipelines. Artifacts can mimic biological features and confound automated detection.

\section{Supporting Tools}
\label{sec:tools}

\textbf{Web-interface:}
The CryoET Data Portal features an interactive web interface powered by Neuroglancer \cite{maitin_shepard_neuroglancer_2021}, a browser-based visualization tool that allows users to explore tomograms and their annotations in 3D. Each tomogram opens with preloaded annotations, enabling users to inspect orthogonal 2D slices and arbitrary 3D cross-sections. 
The interface supports intuitive navigation through gestures and hotkeys, while the control panel offers rendering settings such as opacity, contrast, and layer visibility for segmentations, point annotations, and 3D volumes. 
Tomograms and annotations can also be downloaded in various formats, with detailed metadata accessible through contextual panels.
Access is cloud-optimized through the use of Neuroglancer precomputed formats, OME-Zarr for scalable volumetric data access, and integration with AWS S3 and the Portal API for efficient remote downloads.
Overall, the CryoET Data Portal provides an integrated and user-friendly environment for interactively visualizing and exploring cryoET datasets directly in the browser.

{\bfseries CryoET Data Portal API: }
The CryoET Data Portal provides a Python client for interacting with its GraphQL-based API, enabling users to programmatically search, retrieve, and download datasets, tomograms, annotations, and associated metadata. The API supports flexible querying via find and \texttt{get\_by\_id} methods across core data classes (e.g., Dataset, Run, Tomogram, Annotation), with built-in support for logical and pattern-matching operators. Data objects can be converted to dictionaries or downloaded directly in formats such as MRC and OME-Zarr. The client is installable via pip and designed for use on Linux or macOS systems. The API facilitates reproducible workflows, large-scale data access, and integration into ML pipelines.

{\bfseries Copick: }
copick is an open-source, storage-agnostic Python API and tool suite for collaborative annotation and analysis of cryo-electron tomography data~\cite{copick}. It provides unified access to tomograms, segmentations, meshes, point annotations, and feature volumes regardless of whether they live on local filesystems, HPC clusters, cloud storage, or public repositories such as the CryoET Data Portal. Projects can be partitioned into shared read-only and user-specific mutable layers, allowing multiple researchers to annotate the same data without conflicts, and a multi-resolution storage layout supports responsive visualization even of remote data. Plugins for ChimeraX and napari allow human-in-the-loop particle picking, segmentation, and inspection of machine-learning outputs, while companion libraries provide command-line operations for annotation processing and PyTorch integration for model training. A Model Context Protocol (MCP) server further allows LLM agents to compose curation pipelines from natural-language descriptions of biological constraints. To support reproducible dataset publication, copick implements an mlcroissant backend that exports projects as standards-compliant Croissant manifests with integrity hashes, and train/val/test split declarations, so published datasets can be consumed either self-contained or as a remote read-only reference paired with a local annotation overlay.

\end{document}